\documentclass[aps,prb,reprint,groupedaddress,showpacs,nofootinbib]{revtex4-1}
\usepackage{caption}
\usepackage{dcolumn}
\usepackage{bm}

\newcommand{\lsim}{\raisebox{-0.7ex}{$\;\stackrel{\textstyle <}{\textstyle\sim}\;$}}
\input{epsf}

\begin{document}
\title{Conductivity of a two-dimensional HgTe layer near the critical width:\\ The role of developed edge states network and random mixture of $p$- and $n$-domains}

\author{M.M. Mahmoodian}
\email{mahmood@isp.nsc.ru}
\affiliation{Rzhanov Institute of Semiconductor Physics, Siberian Branch, Russian Academy of Sciences, Novosibirsk, 630090 Russia}
\affiliation{Novosibirsk State University, Novosibirsk, 630090 Russia}

\author{M.V. Entin}
\email{entin@isp.nsc.ru}
\affiliation{Rzhanov Institute of Semiconductor Physics, Siberian Branch, Russian Academy of Sciences, Novosibirsk, 630090 Russia}
\affiliation{Novosibirsk State University, Novosibirsk, 630090 Russia}

\date{\today}

\begin{abstract}
The conductivity of a two-dimensional HgTe quantum well with a width $\sim$6.3~nm, close to the transition from ordinary to topological insulating phases, is studied. The Fermi level is supposed to get to the overall energy gap. The consideration is based on the percolation theory. We have found that the width fluctuations convert the system to a random mixture of domains with positive and negative energy gaps with internal edge states formed near zero gap lines. In the case with no potential fluctuations, the conductance of a finite sample is provided by a random edge states network. The zero-temperature conductivity of an infinite sample is determined by the free motion of electrons along the zero-gap lines and tunneling between them.

The conductance of a single $p$-$n$ junction, which is crossed by the edge state, is found. The result is applied to the situation when potential fluctuations transform the system to a mixture of $p$- and $n$-domains. It is stated that the tunneling across $p$-$n$ junctions forbids the low-temperature conductivity of a random system, but the latter is restored due to the random edge states crossing the junctions.
\end{abstract}


\maketitle
\section{Introduction}
Topological insulators (TIs) have attracted a great deal of attention in the past decade \cite{hasan,qi,thou,wen,kane1,kane2,fu1,moore,roy,bhz,fu2,konig,hsieh,xia,zhang} (more references can be found in \cite{hasan,qi}). Like ordinary insulators (OIs), TIs have an energy gap between occupied valence and empty conduction bands. However, unlike ordinary insulators, they have a negative energy gap sign. Due to topological reasons, this inevitably leads to conductive TI borders. The edge state energies cover the entire energy gap.

The idea of TIs is similar to the classification of electron states in the quantum Hall effect \cite{thou,wen} based on topological order. The edge states in the quantum Hall effect are characterized by the topological phase, which produces the gapless boundary modes that are insensitive to smooth changes in material parameters. In TIs, the role of the magnetic field is transferred to the spin-orbit interaction.

The most widely known representative of a TI is a two-dimensional (2D) HgTe quantum layer \cite{kvon,bhz,vp,qi,shen}. In two dimensions, the backscattering processes in the edge states are strongly forbidden by the time reversibility. In such a case, the electron transport should be one-dimensional, spin-conserving and non-local. However, experimental observations show that many aspects of this picture contradict the ideal picture \cite{kvon}. In particular, this concerns the absence of 2D transport and the backscattering on the edge states.

The edge states in a 2D TI are often considered based on the six-band Bernevig-Hughes-Zhang \cite{bhz} model with zero boundary conditions on the external border. Less known is the minimal Dirac-like two-band model by Volkov-Pankratov (VP) \cite{vp}, which was first invented for the 3D case and was later used for the 2D case \cite{ent-mah-mag,ent-brag1,ent-brag2,mag-ent,mah-mag-ent}.

In the VP model, the gap in the system continuously changes its value passing through zero, and it provides electron states localized near the zero gap line (ZGL). Different aspects of this model have been studied recently. The VP model specificity is the linearity of the edge-state spectrum \cite{ent-mah-mag}, which is symmetric around the gap center. The linearity leads to the suppression of the electron-electron interaction \cite{ent-brag1,ent-brag2}. The edge states on the curved edges were also studied \cite{mag-ent}. It was found that the microwave absorption in the insulating phase is a result of the transitions from the edge state to 2D states or between the edge states with the opposite direction of motion with virtual participation of the 2D states \cite{mah-mag-ent}.

\begin{figure}[ht]
\centerline{\epsfysize=2.5cm\epsfbox{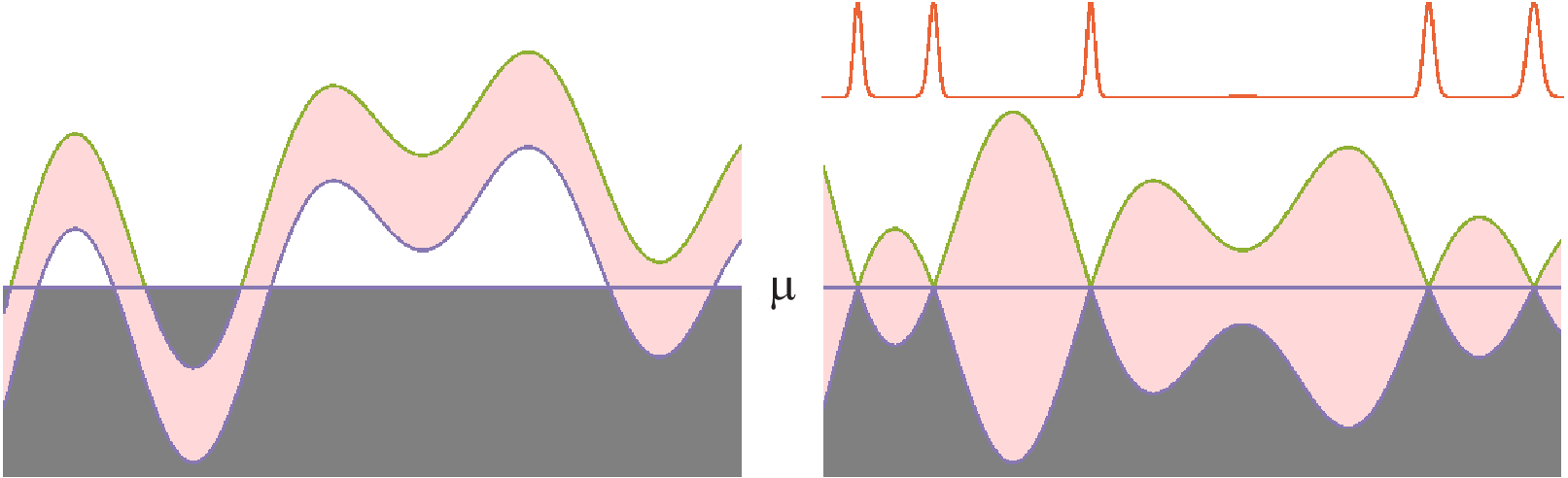}}
\caption{Left: The narrow-gap semiconductor with developed potential fluctuations. The energy gap is marked in light-red. The Fermi level is located in the mean gap center. If the hybridization of states in electron and hole lakes is weak, the system is insulating. The strong hybridization of these states converts the system to a metal. Right: The narrow-gap semiconductor with developed gap fluctuations. The edge states appear near ZGLs. The electron density at the edge states is plotted at the top of the figure.}\label{fig1}
\end{figure}

The 2D TI with potential disorder was investigated in \cite{girsch}. Unlike \cite{girsch}, we study the system with a disordered energy gap. The difference between the approaches in the present paper and \cite{girsch} is schematically demonstrated in Fig.~\ref{fig1}.

Note that the edge transport model in the TI has some similarity to the adiabatic transport in the quantum Hall effect (QHE) with a random potential \cite{bas-mag-ent,iordansky}. In the last problem electrons in a strong magnetic field move along the lines of constant potential, while, in the random-gap TI, the edge states appear with propagating electrons along the ZGLs.

\begin{figure}[ht]
\centerline{\epsfysize=5cm\epsfbox{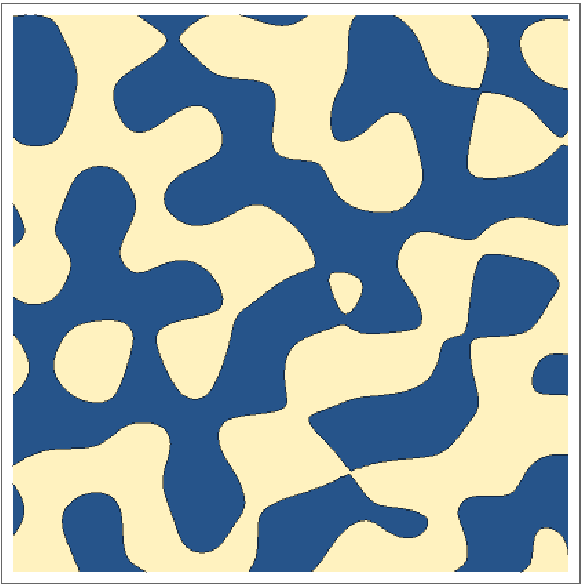}}
\caption{Relief of the random function $\Delta({\bf r})$ near the critical width. The domains of $\Delta({\bf r})<0$ are blue and those of $\Delta({\bf r})>0$ are white.}\label{fig2}
\end{figure}

\begin{figure}[ht]
\centerline{\epsfysize=5cm\epsfbox{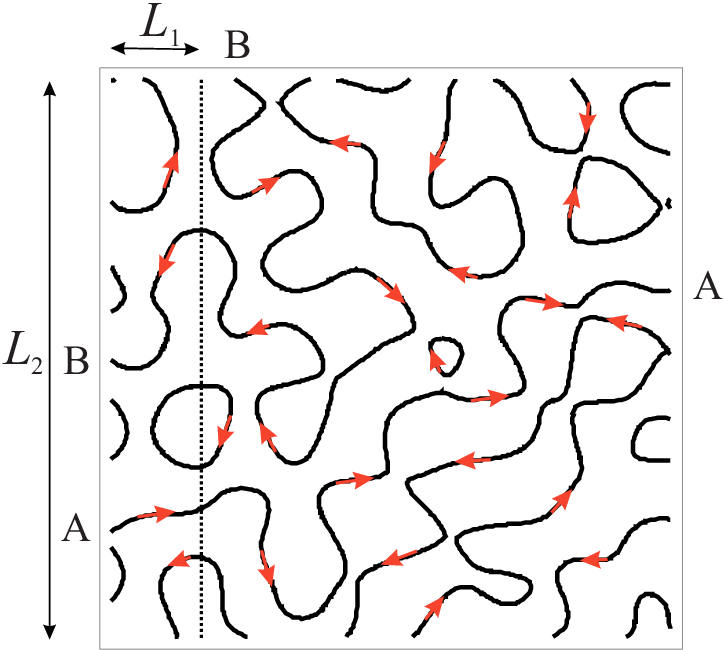}}
\caption{(Color online) The lines of $\Delta({\bf r})=0$, corresponding to Fig.~(\ref{fig2}), along which the edge states are located. The electron motion direction at a fixed spin projection is shown by arrows. In a square sample there is one ZGL AA, percolating in the $x$-direction, and no percolation in the $y$-direction. In the sample elongated in the $y$-direction (separated by a dashed line from the square), there are two edge lines crossing the sample in the $x$-direction (AA and BB) and no such ZGLs in the $y$-direction.}\label{fig3}
\end{figure}

The purpose of the present paper is to study 2D low-temperature stationary electron conductivity in a 2D TI. Our specific interest is focused on the system with a near-critical thickness $w\sim$~6.3~nm, where the energy gap changes its sign. In this situation the gap relief randomness leads to the edge states network formation [see Figs.~\ref{fig1} (right), \ref{fig2}, and \ref{fig3}].

\section{Problem formulation}

The remainder of this paper is organized as follows. First, we shall formulate the random VP model (Sec. III), which is utilized here for the internal edge states consideration. Then, we shall study the conductance of the $p$-$n$ junction with a crossing edge state. This state produces a channel short-circuiting the insulating $p$-$n$ junction (Sec. IV). Then, the conductance of the finite sample with the fluctuating gap sign and without potential fluctuations will be studied (Sec. V). We shall describe the edge states in the framework of percolation theory (Sub. VA). The random edge states network is formed in such a system. The problem can be reformulated as a motion with a constant velocity of particles along ZGLs of a random function $\Delta({\bf r})$. In such an approach, the problem of electron transport converts to the study of geometrical properties of ZGLs (Sub. VB).

Near the sample threshold width, the conductance experiences strong fluctuations and depends on the sample shape. We shall find the conductance of wide or long samples. The percolation consideration shows that the conductivity of an infinite sample tends to zero, unless the interedge transitions are taken into account (Sub. VB). The problem solution is given by taking into account the finiteness of the edge state width that provides the interedge transitions resulting in the 2D conductivity at zero temperature (Sec. VI).

The other studied approach is the conductivity near the conduction- and valence-band thresholds in the presence of an energy gap and potential fluctuations (Sec. VII).

Furthermore, we consider the case in which the system, in the presence of strong potential fluctuations, converts to a mixture of $p$- and $n$-domains (Sec. VIII). These domains have the same proportions at the charge neutrality point (CNP). If the Fermi level deviates from the CNP, the electron or hole liquids form connected domains covering the entire sample. In this case, the current flows through the $n$- or $p$-domain, while the opposite phase is insulated from the current by $p$-$n$ junctions. The conductivity vanishes near the thresholds of the connectivity of these domains.

In the CNP the $p$-$n$ junctions block the overall conductivity. However, the edge states can produce short-circuites of $p$-$n$ junctions that restore the zero-temperature conductivity. This situation will be considered using the conductance of a single $p$-$n$ junction with a crossing edge state. The channel conductance will be found using a single-mode approximation (as half of the conductance quantum). Then the fractal geometry of the edges will be applied to estimate the characteristic conductances and 2D conductivity. This picture is valid if the $p$-$n$ junction widths are large enough to block the tunneling. In the next section we shall consider the model problem of a single planar $p$-$n$ junction, which is crossed by a short-circuiting edge state. All the obtained results will be reviewed in Sec. IX.

The experiment of \cite{kvon} shows 2D TI resistivity growth with the critical width near the neutrality point at a low temperature. The resistivity maximum reaches the value of some resistivity quanta, while, apart from this point, the resistivity is much lower. This experimental observation is important for a suitable theory.

\section{Random Volkov-Pankratov model}
A basic assumption of this paper is that the HgTe layer experiences the transition between ordinary (with a positive gap) and inverted (with a negative gap) insulating phases (OI and TI), accordingly, when the mean well width $\overline{w}$ changes from zero to infinity at $\overline{w}=w_0=$~6.3~nm. We suppose that the energy gap $2\Delta$ has the linear dependence on the well width $w({\bf r})$ [${\bf r}=(x,y)$]: $\Delta(w)=\alpha(w-w_0)$, $\alpha=\partial\Delta(w)/\partial w|_{w=w_0}$. Using the data from \cite{qi}, we obtain $\alpha\approx-8.75~\mbox{meV}~\mbox{nm}^{-1}$.

Near the critical width $w_0$, the inevitable fluctuation of $w({\bf r})$ leads to the separation of a sample to OI and TI domains. The borders between them, where the gap $2\Delta({\bf r})=0$, should form the edge states \cite{ent-mah-mag,ent-brag1,ent-brag2,mag-ent,mah-mag-ent,mah-ent}.

Electrons are described by the 2D Volkov-Pankratov Hamiltonian (VP)
\begin{equation}\label{VP}
H=\left(
    \begin{array}{cc}
      V({\bf r})+\Delta({\bf r}) & v\bm{\sigma}{\bf p} \\
       v\bm{\sigma}{\bf p} & V({\bf r})-\Delta({\bf r}) \\
    \end{array}
  \right),
\end{equation}
where $V({\bf r})$ is the potential, ${\bf p}=(p_x,p_y) $ is the 2D momentum operator, $\sigma_i$ are the Pauli matrices.

If $V({\bf r})$ and $\Delta({\bf r})$ are constant, the Hamiltonian (\ref{VP}) has the energy spectrum $\pm\sqrt{\Delta^2+v^2p^2}$ with the gap $2\Delta$.

If the system has one straight edge, $V({\bf r})=0$, $\Delta({\bf r})=\Delta(y)$, $\Delta(y<0)>0$, $\Delta(y>0)<0$, the VP Hamiltonian yields the edge states with a wave function
\begin{eqnarray}\label{wf}
\Psi_\sigma=
\left(
      \begin{array}{c}
            1+\sigma \\
            1-\sigma \\
            \sigma-1 \\
            1+\sigma \\
      \end{array}
\right)\exp\left(ipx+\frac{\sigma}{v}\int\limits_0^y\Delta(y')dy'\right),
\end{eqnarray}
where $\sigma=\pm1$ is a spin quantum number (here and in what follows, $\hbar=1$; in the final equations we restore the dimensionalities).

The edge states have a linear spectrum:
\begin{equation}\label{eps}
\epsilon=\sigma vp.
\end{equation}
Here $v$ plays the role of the edge-state electron velocity. Below we deal with a smooth dependence of $\Delta({\bf r})$ on ${\bf r}$. In a particular case of $\Delta(y)$, assuming that $\Delta(0)=0$ and expanding $\Delta(y)\approx-\varepsilon y$, $\varepsilon=-d\Delta(y)/dy|_{y=0}$, we get to $\psi_\sigma\propto\exp(-\varepsilon y^2/2v)$ that, at $\varepsilon>0$, yields the wave function localized near $y=0$ with the edge-state width $l_{ed}=\sqrt{8v/\varepsilon}$. The edge states spectrum overlaps the bandgap of the infinite system.

The paper deals with $V({\bf r})$ and $\Delta({\bf r})$ randomly depending on both coordinates. We will study the edge states in a quasiclassical system, where the characteristic planar sizes of potential $b$ and gap $a$ are large, as compared with $l_{ed}$ [for $\Delta({\bf r})$ depending on both coordinates $l_{ed}\sim\sqrt{8v/|\nabla\Delta|}$]. The functions $V({\bf r})\pm\Delta({\bf r})$ represent the conduction band bottom and the valence band top, correspondingly. Below we assume that $V({\bf r})$ and $\Delta({\bf r})$ are independent random functions with Gaussian distributions. The first one results from the random distribution of charge impurities, and the second one results from quantum well width fluctuations.

The quantity $\Delta({\bf r})$ is characterized by its mean value $\overline{\Delta({\bf r})}=\alpha(\overline{w}-w_0)$ and mean-squared fluctuations $\Delta_2\equiv\sqrt{\overline{\delta\Delta^2}}=|\alpha|w_2\equiv|\alpha|\sqrt{\overline{w^2}-\overline{w}^2}$; the overline stands for the spatial mean. The mean-spatial sizes of $\Delta({\bf r})$ and $V({\bf r})$, $a$ and $b$, are: $a^2=w_2^2/\overline{(\nabla w)^2}$, $b^2=\overline{\delta V^2}/\overline{(\nabla V)^2}$.

In the quasiclassical approximation, the ZGLs $\Delta({\bf r})=0$ can be considered as locally straight. In such case one can apply Eq.~(\ref{wf}).

An additional simplification can be done if to assume that wavelength $\hbar/p$ is less than the ZGL characteristic length. This permits a quasiclassical description of the electron motion along the edges. Such particles have the Hamiltonian function $\sigma vp+V({\bf r}(u))$, where $u$ is a coordinate along the ZGL and $p$ is a conjugated momentum. In accordance with this Hamiltonian, independently from the potential, electrons move along the edge with constant velocity $\sigma v$. If necessary, the electron motion along the edge can be quantized \cite{ent-brag1}.

\section{Conductance of a planar $p$-$n$ junction with a crossing edge state}
\begin{figure}[ht]
\centerline{\epsfysize=5cm\epsfbox{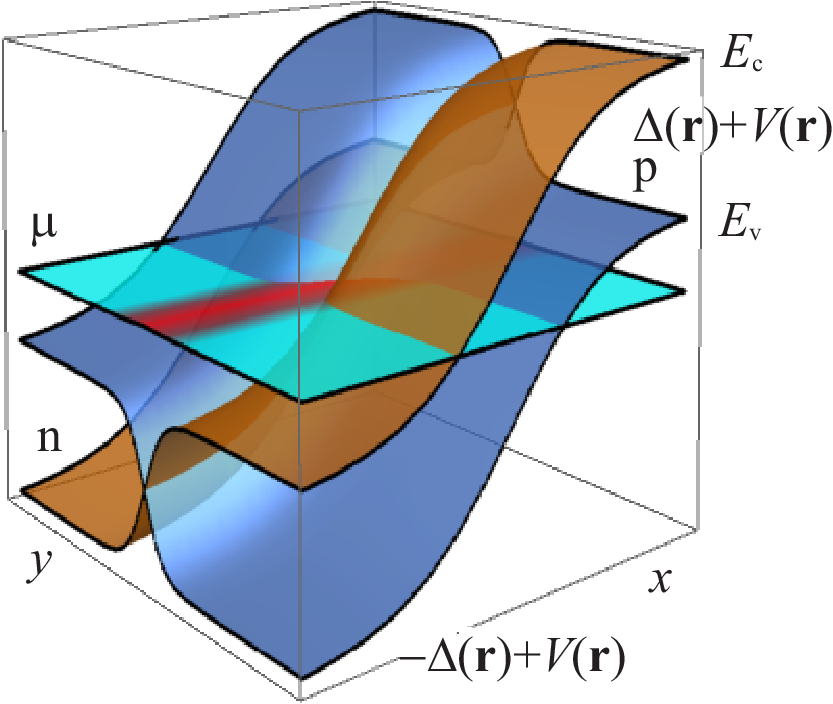}}
\caption{The $p$-$n$ junction with an alternating gap. The blue (upper in the right plot section) and brown (lower) surfaces depict the position of the bands extrema. The central horizontal plane represents the Fermi level; the color on it corresponds to the edge state density.}\label{fig4}
\end{figure}

Consider a potential in the planar $p$-$n$ junction with the impurity charge density distribution as $en\tanh(x/b)$, where $e$ is the electron charge, and $n$ is the carrier density at infinity. Planar charge carriers screen this distribution. That determines the potential across the $p$-$n$ junction (see Fig.~\ref{fig4}).
The potential caused by the 2D charge distribution is
\begin{equation}\label{phi}
V=\int d^2r'\frac{e(n({\bf r}')-n_e({\bf r}'))}{\kappa|{\bf r}-{\bf r}'|},
\end{equation}
where, $\kappa$ is the dielectric constant. In the other form,
\begin{equation}\label{phi1}
V=\frac{1}{\kappa}\int d x'e(n(x')-n_e(x))\ln|x-x'|.
\end{equation}
If the 2D screening length is small enough, the $p$ and $n$ domains are neutral, and the potential is formed by the charge in the insulating domain $-l_d/2<x<l_d/2$. This gives the potential difference
\begin{equation}\label{phi2}
V=\frac{1}{\kappa}\int\limits_{-l_d/2}^{l_d/2}dx'en(x')\ln|x-x'|
\end{equation}
and the boundary condition for the insulating domain width
\begin{equation}\label{phi3}
2\Delta_0=\frac{e^2 n}{\kappa}\int\limits_{-l_d/2}^{l_d/2}dx\tanh\left(\frac{x}{b}\right)\ln|x|.
\end{equation}
Here $2\Delta_0$ is the gap on infinity. If $b\ll l_d$,
$$l_d\approx\frac{2\Delta_0\kappa}{ne^2\ln(\Delta_0\kappa/ne^2b)}.$$

The characteristic decrement for the tunneling is $\Delta_0/v$, and if $l_d\Delta_0/v\gg 1$, the $p$-$n$ junction is impermeable. The parameter $l_d\Delta_0/v$ can be rewritten as $l_d\Delta_0/v\sim (\Delta_0/E_F)(v_F/e^2)(v/v_F)$, where $E_F$ and $v_F$ are the Fermi energy and velocity. All 3 factors in the right part are greater than 1: $(\Delta_0/E_F)\gg 1$ if electrons (holes) fill the extrema of the spectrum, which is an ordinary situation, $v_F/e^2\gg 1$ for a weakly interacting electronic liquid, and $v_F>v$ always.

Then we consider the edge state that crosses the straight $p$-$n$ junction. Let the potential profile have the form of Eq.~(\ref{phi1}), let the gap be $2\Delta(y)=2\Delta_0 \tanh(y/a)$, let the potential $V(x)=\Delta_0\tanh(x/l_d)$, and let the parameters satisfy inequalities $\Delta_0-V_0<\mu<V_0-\Delta_0$ (see Fig.~\ref{fig4}). The gap goes to constants $\pm \Delta_0$ at $y\to\mp\infty$. The potential goes to $\pm \Delta_0$ at $x\to\mp\infty$. In such system, the $x\to\mp\infty$ domains belong to $n$ and $p$, correspondingly. In that case the electron gas in $p$- and $n$-domains is degenerate. Apart from the $y=0$ line, the conductivity across the edge at a low temperature vanishes. The line $y=0$, where $\Delta(y)=0$, crosses the $p$-$n$ junction. Only this place is responsible for the $p$-$n$ junction conductance.

Let there be a single edge state along this line. Independently from the potential, the edge-state conductance is $e^2/h$, where $h=2\pi\hbar$. Hence, this value will determine the total conductance of the $p$-$n$ junction $\Sigma_{junc}$. The edge $y=0$ plays the role of the $p$-$n$ junction short circuit.

Now, consider the case when width $a$ is large enough. Then, electrons are not quantized by the conducting channel width. Considering the tunneling in any point $y$ in an independent manner we can find the probability of tunneling across the junction at a coordinate $y$
$$P_t(y)=\exp\left(-\frac{2}{v}\int\sqrt{\Delta^2-V^2(x)}dx\right)=e^{-\frac{2\pi l_d}{\hbar v}\Delta_0|\tanh\frac{y}{a}|}.$$

The ballistic motion channels are determined by $P_t\sim 1$. This yields the characteristic width $l_t\sim va/l_d\Delta_0$. The number of quantum channels for ballistic tunneling is determined by this width ratio to the edge state width: $l_t/l_d=\sqrt{va/8\Delta_0}$. The $p$-$n$ junction conductance $\Sigma$ is $l_t/l_d\cdot e^2/h$. This approach is valid if $l_t\gg l_d$; otherwise, the approach with a single channel holds. Summarizing this, we have
\begin{equation}\label{Sigma0}
\Sigma_{junc}= \frac{e^2}{h} f\left(\frac{l_t}{l_d}\right),\end{equation}
where $f(0)=1$, $f(x)=x$ at $x\to \infty$.

Note that the conductance does not change if to apply a finite voltage to the $p$-$n$ junction (less than $2\Delta_0$).

Hence, the conductance across the $p$-$n$ junction is determined by the edge states crossing it. The results of the present section are applicable to the case when the $p$-$n$ junction width is less than a characteristic spatial size of gap $a$ or has the same order.

The $p$-$n$ barriers can block the overall conductivity in a system with a random potential. This conductivity is restored by the edge-states inclusion.

\section{Edge states conductance without potential fluctuations}
\subsection{Proximity to the percolation threshold}

As the properties of our system are determined by quantum well width fluctuations, we should introduce their distribution and ruling parameters. The proximities to the percolation threshold of OI or TI phases are described by the portions of the area belonging to the TI, or OI, $\xi$ and $1-\xi$, correspondingly. The proximity of these parameters $\xi-\xi_c$ to the threshold  $\xi_c$ is a dimensionless quantity that determines all critical exponents. We shall set $\xi_c=1/2$ in accordance with the 2D site percolation model.

For the Gaussian distribution of widths, the corresponding probability is
\begin{eqnarray}\label{xidelta}
\xi=\frac{1}{\sqrt{2\pi}w_2}\int\limits_{w_0}^{\infty}e^{-\frac{\left(w-\overline{w}\right)^2}{2w_2^2}}dw=
\frac12\mbox{erfc}\frac{w_0-\overline{w}}{\sqrt{2}w_2},
\end{eqnarray}
where $\mbox{erfc}(x)$ is a complementary error function.

At the percolation threshold, if $w_0-\overline{w}\ll\sqrt{2}w_2$, $\xi-1/2=(\overline{w}-w_0)/\sqrt{2\pi}w_2$.

\subsection{Quantized conductance}

First, we should specify what we mean by the word ''sample''. The source and drain contacts are metallic. The edge state electrons coming into these contacts are quickly mixed with the electron sea due to elastic scattering. The simplest is the case of a sample without side boundaries. For example, one can consider the ring sample (the ''Corbino'' disk) with metallic inner and outer parts. Another example is an infinite strip across the current direction.

The other situation is with the lateral faces of the rectangular sample. One can not use the limitation due to the potential: edge-state electrons freely come through this $p$-$n$ junction (see previous section). Instead of this, the limitation of the electron motion in the side direction can be done by the contact with a large-gap OI situated outside the sample. However, this procedure inevitably produces the edge states on the side border.

We shall assume that the sample sizes exceed $a$. Let the Fermi level in a ring sample fall inside the forbidden band. At the same time, fluctuations in the gap sign create the edge states on ZGLs. Depending on the percolation threshold proximity, the ZGLs turn out to be closed at $|\xi-\xi_c|\sim1/2$, or open lines appear at $|\xi-\xi_c|\ll 1/2$. In the first case, the conductance of a large sample vanishes, with the exception of the external edges contribution. In the case $|\xi-\xi_c|\ll 1/2$, the conductance appears at zero temperature (see Fig.~\ref{fig3}).

If they also exceed the correlation length, the edge channels at $\xi\neq\xi_c$ are the close lines that do not cross all of the sample. If $\xi\rightarrow\xi_c$, the ZGLs begin to cross the sample; that provides the conductance $\Sigma=N_0e^2/h$, where $2N_0$ is the number of the lines $\Delta({\bf r})=0$ crossing the sample in the field direction (radius for the ring case), and $2e^2/h$ is the conductance quantum. As the ZGLs has no branching, the presence of conductance in the radial direction means no conductance in the angular direction and {\it vice versa}. So, the conductance fluctuates between 0 and a value of the order of $e^2/h$.

The conductance along the external edges $e^2/h$ appears in a square sample. Besides, the additional inner ZGLs crossing the sample in the field direction can exist. So, $\Sigma=(N_0+1)e^2/h$.

\subsection{Consideration based on the percolation theory}

Here we shall consider a finite rectangular sample based on the percolation theory. We shall start from a square sample. Assume that the ZGL is absolutely random and starts from the right edge of a square. Then it has approximately equal opportunities to cross any other square edges, and the probability of crossing the sample in the field direction is 1/4.

In a sample elongated in the applied voltage direction, with width $L_2$ and length $L_1\gg L_2$, the percolation probability should qualitatively be a product of the probabilities of percolation through the square samples from which it is composed (Fig.~\ref{fig3}). So, for an elongated sample, the percolation probability should be proportional to $1/4^{L_1/L_2}$. Hence, the mean conductance of such a sample decays exponentially with its length.

The situation is different for a wide sample ($L_2>L_1$). Blocks $L_1\times L_1$ are independent of each other. Hence, the conductances are added in a parallel manner.

The previous consideration was too qualitative. To be more accurate, let us include the percolation theory. For conductance finiteness, ZGLs should connect the opposite sides of the sample $x=0$ and $x=L_1$. The probability of one of the phases in connecting the borders can be found using the correlation function $G(r)$ \cite{shkl-efros}. In the percolation threshold vicinity
$$G(r)=\left(\frac{a}{r}\right)^\eta e^{-r/L_c},$$
where $L_c=a|\xi-\xi_c|^{-\nu}$, $\eta=2-\frac{\gamma}{\nu}\approx 0.22$, $\gamma\approx 2.38$, and $\nu\approx 1.34$. Near the percolation threshold, the probability, for the edge, of crossing the sample in points ${\bf r}_1=(0,y_1)$, ${\bf r}_2=(L_1,y_2)$, $0<y_1$ and $y_2<L_2$, has the same order as the probability of ${\bf r}_1$ and ${\bf r}_2$ points to belong to the same cluster, namely $G(|{\bf r}_1-{\bf r}_2|)a^{-2}dy_1dy_2$. The fractal dimension of the large cluster hull is close to 1 (see \cite{voss}). This follows from the fact that, on the threshold, exactly half of the neighboring sites of OI belongs to an OI or a TI subset. So, the probability of that a site belongs to a cluster border has the order of 1; for example, in the site problem on a square lattice, this probability is 15/16). As a result, we can replace the correlation function of edge points by the correlation function of OI or TI sites.

Now, let us consider a $L\times L$ square sample. The probability of percolation along the edge in the $x$-direction is collected from the probability for two points on the sample borders $x=0$ and $x=L$ of belonging to the same cluster $G(|{\bf r}_1-{\bf r}_2|)$ and the probability $\exp{(-Z)}$ that none of the other points of the rectangle borders $\Gamma$ to be connected with the starting point. We should find the number of ZGLs, which cross the opposite edges $x=0$ and $x=L$, with the limitation that these lines remain inside the rectangle. It is the product of the number of points ${\bf r}_1$ and ${\bf r}_2$ belonging to the same cluster
$$N_c=\int\limits_0^{L}\int\limits_0^{L}G(|{\bf r}_1-{\bf r}_2|)a^{-2}dy_1dy_2,$$
where ${\bf r}_1=(0,y_1)$, ${\bf r}_2=(L,y_2)$, and the ZGL probability of not crossing all borders in other points is given by a product
\begin{eqnarray}\label{prob}
\prod\limits_{{\bf r}_2\epsilon\Gamma}\left(1- G(|{\bf r}_1-{\bf r}_2|)a^{-1}dt_2\right)=e^{-Z},\\
Z=\int\limits_0^{L}\int\limits_{{\bf r}_2\epsilon\Gamma}G(|{\bf r}_1-{\bf r}_2|)\frac{dy_1 dt_2}{La}.
\end{eqnarray}
Here $t_2$ is the length along $\Gamma$.

The system conductance $\Sigma(L)$ is the product of the half conductance quantum and the probability of connection of two opposite sample sides:
\begin{eqnarray}\label{Sigma}
\Sigma(L)\sim\frac{e^2}{h}N_c\exp{(-Z)}.
\end{eqnarray}

In the limiting cases at $L\ll L_c$
\begin{eqnarray}\label{Sigma1}
\Sigma(L)=c_1\frac{e^2}{h}\left(\frac{L}{a}\right)^{2-\eta}e^{-c_2\left(\frac{L}{a}\right)^{1-\eta}},
\end{eqnarray}
and at $L\gg L_c$
\begin{eqnarray}\label{Sigma2}
\Sigma(L)=c_3\frac{e^2}{h}\left(\frac{L}{a}\right)^{\frac32-\eta}\left(\frac{L_c}{a}\right)^{\frac12}e^{-\frac{L}{L_c}-c_4\left(\frac{L_c}{a}\right)^{1-\eta}}.
\end{eqnarray}
Here  $c_3=\sqrt{\pi/2}\approx1.25$; at $\eta=0.22$ quantities $c_1\approx0.97$, $c_2\approx4.56$ and $c_4\approx1.19$. If $L\ll L_c$, $\Sigma(L)$, experiences a power-like drop with $L$; if $L\gg L_c$,  $\Sigma(L)$ exponentially drops.

Consider the conductance of a rectangular sample, when $L_2\gg L_1$ or $L_1\gg L_2$. Like the approximate consideration, the conductance of samples with $L_2\gg L_1$ and $\Sigma(L_1,L_2)$ is the sum of squares conductances:
\begin{eqnarray}\label{Sigma3}
\Sigma(L_1,L_2)=\Sigma(L_1)L_2/L_1.
\end{eqnarray}
This is valid in a very wide sample or in a ring sample. In a sample of finite width, one should add $e^2/h$ to this value.

In the case of $L_1\gg L_2$, the percolation probability exponentially decays with the length $L_1$ at $L_1\gg L_c$, while the external edge conductance remains the same. This means that
\begin{eqnarray}\label{Sigma4}
\Sigma(L_1,L_2)=2e^2/h+o\left(\exp{(-L_1/L_c)}\right).
\end{eqnarray}
The factor 2 accounts for the fact that the external edges appear pairwise.

We should emphasize that Eqs.~(\ref{Sigma})-(\ref{Sigma2}) give estimations only. Besides, the conductance is strongly fluctuating for the samples with $L_2\lsim L_1$, while, at $L_2\gg L_1$, the conductance is self-averaging.

There are different reasons for the inaccuracy of the obtained results. First, we replaced the edge sites correlation function with the TI phase correlation function (when the OI phase prevails) or the OI phase in the opposite case. In principle, the behavior of both correlation functions should be similar, but the corresponding exponents can differ. We hope that, in accordance with that mentioned above, this numerical difference is not strong. The other inaccuracy consists in using the percolation correlation functions for the unbounded sample to describe the strip. This inaccuracy manifests itself in the difference of the percolation probability along the strip obtained by the ZGL in non-correlated and correlated models (different probability logarithm dependence on the strip length). Again, this difference is inessential in the sufficiently rough approximation which is used here. This consideration neglects the tunneling between ZGLs, which is considered in the next section.

\section{2D edge conductivity at zero temperature. Finite edge state width}

\begin{figure}[ht]
\centerline{\epsfysize=5cm\epsfbox{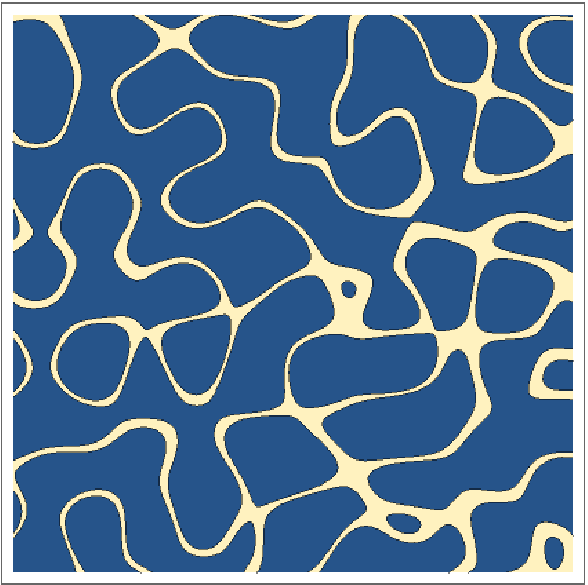}}
\caption{Edge states with a finite width (light-yellow). As compared with Fig.~(\ref{fig3}), the percolation appears in both directions.}\label{fig5}
\end{figure}

Here we discuss the consequences of the edge-state width finiteness. It is obvious that the exaggerated picture of the transport along ZGLs is limited by the edge width. If the width becomes comparable with the distance between different ZGLs or different parts of the same line, the intensive tunneling will destroy this picture. Instead of strictly following ZGLs, an electron can jump from one place to another (see Fig.~\ref{fig5}).

The free jump distance is determined by the edge width $l_{ed}$, which can be estimated as $\sqrt{av/\Delta_0}$. At $\xi\ll 1$ or $1-\xi\ll 1$, we suppose the equality of the densities of internal and edge points of clusters. Interpolating for all $\xi$, we can set the edge points density $N_e=\xi(1-\xi)/a^2$. Neglecting the edge points correlation, one can estimate the threshold $\widetilde{\xi}_c$ when the transport along ZGLs converts to the two-dimensional traveling due to jumps as $\pi N_el_{ed}^2=B_c$. Here $B_c$ is some number calculated in the percolation theory. $B_c$ runs from 3.2 to 4.5 \cite{shkl-efros} in different 2D percolation models. As a result, $\widetilde{\xi}_c(1-\widetilde{\xi}_c)=B_ca\Delta_0/2\pi v$. If $\xi(1-\xi)<\widetilde{\xi}_c(1-\widetilde{\xi}_c)$, the edge traveling prevails; otherwise, two-dimensional jumps occur and that means the 2D delocalization and, in such a case, the conductivity should be determined by the Drude-like expression.

Note that we neglected the quantization of electron motion along edges. The longitudinal quantization results in a minimal energy needed to jump from one close edge state to another with a characteristic quantum of the order of $\xi(1-\xi)v\hbar/a$. This quantum should be less than the characteristic hopping amplitude $\Delta \exp{(-(\Delta/\hbar va)/N_e)}$, where the characteristic hopping distance $l_{hop}=1/\sqrt{N_e}$. Obviously, this is impossible and the quantization is inessential.

The developed ZGLs near the threshold $\xi\to\xi_c$ are long. There is a finite probability that, somewhere, such lines will approach the other ZGL to a distance comparable with the edge-state width. That results in the electron possibility of jumping from one ZGL to another, and the 2D diffusion will be established.

We shall find the diffusion coefficient and the conductivity based on the model of random $\Delta({\bf r})$ with the Gaussian distribution. Traveling along a ZGL, an electron meets a different relief of random $\Delta({\bf r})$. The quasiclassical edge-state width depends on the coordinate along the edge as $l_{ed}\sim \sqrt{8v/|\nabla\Delta({\bf r})|}$. (The quasiclassical approach is valid if $a\gg l_{ed}$.)

The edge-to-edge transition occurs in places where two ZGLs are closer to each other than $l_{ed}$. Let us introduce two curvilinear coordinates: $t$ - along ZGL, $\Delta({\bf r})=0$; and $\rho$ - across it. In these coordinates, $\Delta(t,\rho=0)=0$. Besides, in the vicinity of point $t_0$ of a minimal distance between two ZGLs, one can expand $\Delta(t,\rho)$ as
\begin{eqnarray}\label{eqq}\nonumber
\Delta(t,\rho)=\partial_\rho \Delta(t_0,0)\rho+\partial_\rho^2\Delta(t_0,0)\frac{\rho^2}{2}+\\
+\partial^3_{t,t,\rho}\Delta(t_0,0)(t-t_0)^2\frac{\rho}{2}.
\end{eqnarray}

Eq.~(\ref{eqq}) takes into account the need for having two close solutions for the ZGLs:
\begin{eqnarray}\label{eqq1}\nonumber
\rho=0,~~~ \rho=\rho_0 +a_1(t-t_0)^2,\\ \nonumber
\rho_0=-2\frac{\partial_\rho \Delta(t_0,0)}{\partial_\rho^2\Delta(t_0,0)},~~~ a_1=-\frac{\partial^3_{t,t,\rho}\Delta(t_0,0)}{\partial_\rho^2\Delta(t_0,0)}.
\end{eqnarray}

The quantity $\rho_0$ is the minimal distance between the ZGLs reached at point $t_0$. The condition for a jump between edges is $\rho_0<l_{ed}$. In other terms,
\begin{eqnarray}\label{cond}
2\hbar v(\partial_\rho^2\Delta(t_0,0))^2-|\partial_\rho \Delta(t_0,0)|^3>0.
\end{eqnarray}
Eq.~(\ref{cond}) can be estimated as $\hbar v\Delta^2/a^4>\Delta^3/a^3$. Let us divide the ZGL into cuts of minimal length $a$. The portion of configurations, when Eq.~(\ref{cond}) is valid, is determined by the ratio of value $\Delta$ from this inequality to the mean fluctuation of $\Delta$. Hence, the portion of cuts, where the jumps can occur, is $\hbar v/a\Delta_2$ and the mean free path for a jump is $l_p=a^2\Delta_2/\hbar v$.

Now we should express the Cartesian distance of travel $L$ via $l_p$. This relation is given by the hull fractal dimension: $l_p=a(L/a)^{D_h}$, $D_h\approx 1.74$ \cite{voss}.

Thus, an electron randomly shifts at distance $L$ at the mean free time $l_p/v$. This yields the 2D conductivity $\sigma$ expressed via the diffusion coefficient $D=L^2v/l_p\sim av\left(\Delta_2a/\hbar v\right)^{2/D_h-1}$. The diffusion coefficient is connected with the conductivity and the density of edge states $g$. For a linear spectrum, $g$ does not depend on the energy and is determined by the number of edge nodes per unit area $1/2a$:
\begin{equation}\label{cond2}
\sigma=\frac{e^2D}{2\pi\hbar va}\sim\frac{e^2}{h}\left(\frac{\Delta_2a}{\hbar v}\right)^{\frac{2}{D_h}-1}.
\end{equation}
That explains how the 2D diffusion and 2D conductivity appear. The power $2/D_h-1\approx 0.15$ is small, but positive; hence, the conductivity slightly exceeds the conductance quantum.

The threshold for the appearance of 2D conductivity is determined by the requirement that the gap fluctuations correlation length $L_c$ exceeds $L=a(l_p/a)^{1/D_h}$: $L_c(\xi_c')=a|\xi_c'-\xi_c|^{-\nu}\sim L$, or $|\xi_c'-\xi_c|\sim\left(\hbar v/a\Delta_2\right)^{1/\nu D_h}=\left(\hbar v/a|\alpha|w_2\right)^{1/\nu D_h}$.

The 2D diffusion picture is realized if $|\xi-\xi_c|<\left(\hbar v/a|\alpha|w_2\right)^{1/\nu D_h}$. Otherwise, the 2D diffusion and conductivity vanish.

Recounting for the HgTe width, we find that the 2D conductivity appears if $|\overline{w}-w_0|<\sqrt{2\pi}w_2\left(\hbar v/a|\alpha|w_2\right)^{1/\nu D_h}$ and, on the contrary, the low-temperature conductivity is absent. This shows the extremely unusual behavior of the conductivity with $w$: it exists in a narrow window of widths near $w_0$. The origin of this phenomenon is almost a collisionless electron propagation along the ZGLs together with the growth of the number of contacts between ZGLs when $\overline{w}\to w_0$.

The 2D conductivity is established if all sample sizes exceed the parameter $a(l_p/a)^{1/D_h}$. If both $L_1$ and $L_2$ are less than this quantity, one should use the results of Section VI.

Note that, in this approximation, the transition does not depend on the Fermi level inside the gap. This is explained by the spectrum linearity because all physical properties of such a system do not depend on the electron energy.

\section{Percolation in the system with gap and potential fluctuations near p- and n-type conductivity thresholds}
In a pure OI or TI, the conductivity appears when the Fermi level comes in electron or hole permitted bands. If the chaos is weak, the band edges smear. The conductivity near thresholds is realized through the electron or hole seas. The inner part of $p$- and $n$-regions has the Drude conductivity $\sigma_{e,h}\sim(e^2/h)E_{F;e,h}\tau_{e,h}$ (for $\delta E_{F;e,h}\tau_{e,h}\gg 1$) or minimum metallic conductivity $0.2e^2/h$ (for $\delta E_{F;e,h}\tau_{e,h}\sim 1$), where $E_{F;e,h}$ are the Fermi energies of electrons and holes, and $\tau_{e,h}$ are their mean free times.

Analogously to Sec. VA, for the Gaussian distributions of potential, one can define the probability $\xi_{e,h}$ for conduction and valence bands:
\begin{eqnarray}\label{xiphi}
\xi_{e,h}=\frac{1}{\sqrt{2\pi\overline{\delta V^2}}}\int\limits_{\overline{\Delta}\mp\mu}^{\infty}e^{-\frac{V^2}{2\overline{\delta V^2}}}dV=
\frac12\mbox{erfc}\frac{\overline{\Delta}\mp\mu}{\sqrt{2\overline{\delta V^2}}},
\end{eqnarray}
where $\mu$ is the chemical potential. If one takes into account $\delta w$, the quantity $\overline{\delta V^2}$ in Eq.~(\ref{xiphi}) should be replaced by the mean-squared fluctuations of the conduction (valence) band edges $\delta E^2=\overline{\delta V^2}+\Delta_2^2$.

The probability $\xi_{e,h}$ determines the n- and p-domains connectivity (in infinite systems) by conditions $\xi_{e,h}>\xi_c=0.5$, correspondingly. It is known \cite{shkl-efros} that, when the Fermi level $\mu$ approaches the bands edge, the conductivity has the power-law behavior $\propto |\xi_{e,h}-\xi_c|^t$ with $t\approx 1.38$. Thus, in the case of $\delta E\tau\sim 1$, we have
\begin{equation}\label{sig0}
\sigma\sim\frac{e^2}{h}\left[(\xi_e-\xi_c)^t\theta(\xi_e-\xi_c)+(\xi_h-\xi_c)^t\theta(\xi_h-\xi_c)\right].
\end{equation}

Eq.~(\ref{sig0}) requires that the band edge smearing be less than gap $2\Delta$.

\begin{figure}[ht]
\centerline{\epsfysize=3.5cm\epsfbox{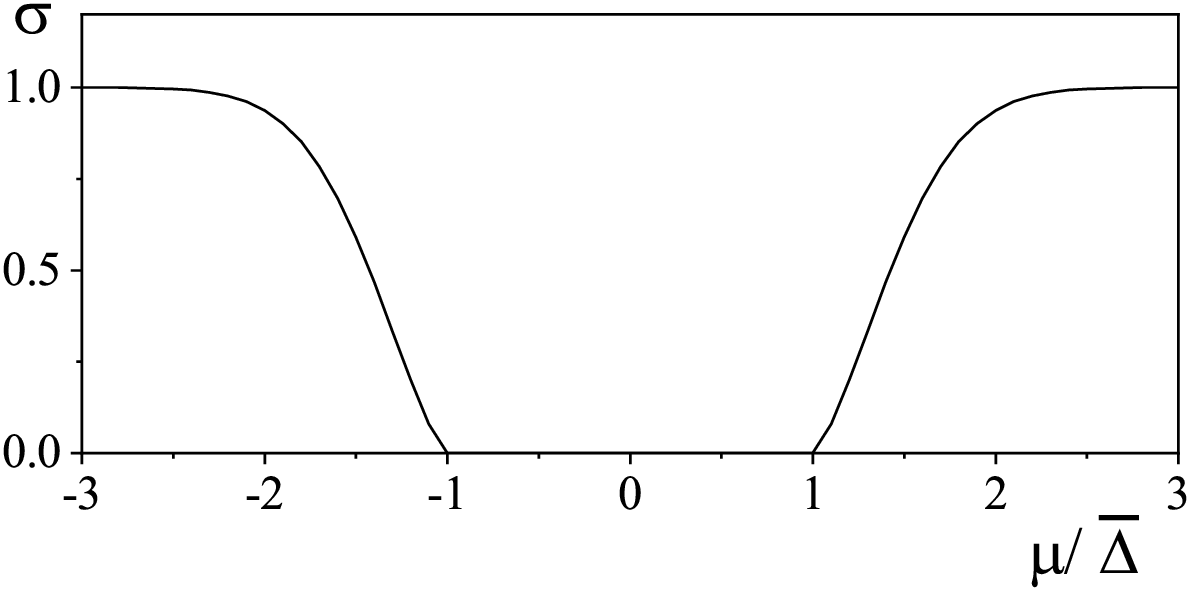}}
\caption{Conductivity (in units of $e^2/h$) dependence on the Fermi level at $\delta E=0.5\overline{\Delta}$, in accordance with Eqs.~(\ref{xiphi}) and (\ref{sig0}).}\label{fig6}
\end{figure}

The conductivity {\it versus} the Fermi level for different average HgTe layer thicknesses is shown in Fig.~\ref{fig6}.

\section{Developed potential fluctuations in the presence of random edge states}
\begin{figure}[ht]
\centerline{\epsfysize=4.5cm\epsfbox{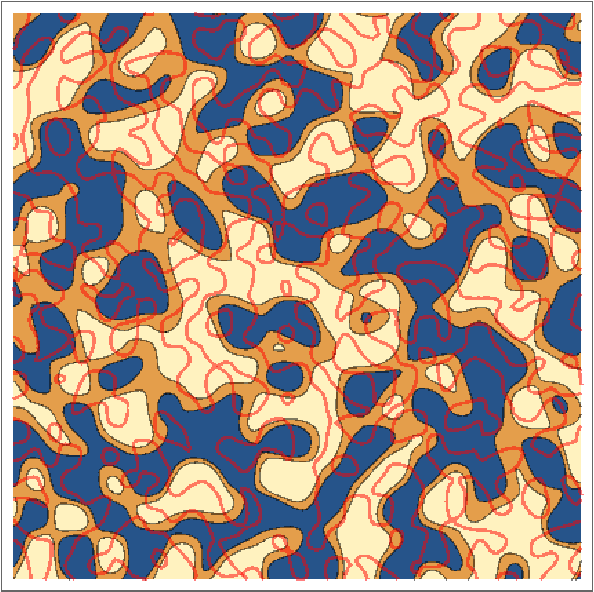}\hspace{0.3cm}\epsfysize=3.5cm\epsfbox{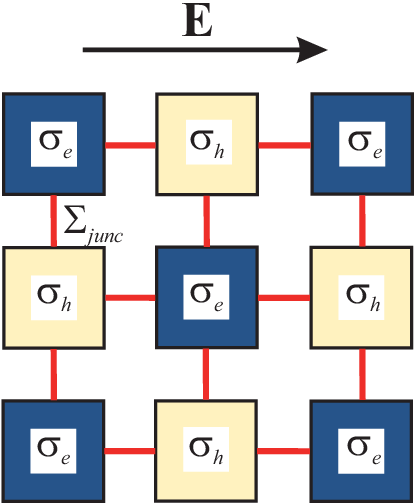}}
\caption{Left: relief of the random potential (blue, light-orange and light-yellow colors correspond to $n$-, $i$- and $p$-domains, correspondingly). Red lines represent the ZGLs short-circuiting $p$-$n$ junctions. Right: equivalent circuit. Squares stand for $p$ and $n$ lakes; lines replace the edge states.}\label{fig7}
\end{figure}

In the absence of edge states, the conductivity of an infinite system inside the bandgap with a fluctuating potential vanishes. Let the potential fluctuations $\delta V$ be large, as compared with the mean gap  $\overline{\Delta}$. Then the system will consist of large $n$ and $p$ metallic lakes (with size $b$) separated by narrow $p$-$n$ junctions. The characteristic $p$-$n$ junction width is $ b\overline{\Delta}/ \delta V \ll b$. The $p$-$n$ junction isolates these regions from each other. The ZGLs $\Delta({\bf r})=0$ cross the $p$-$n$ junctions (see Section III and Fig.~\ref{fig4}) and short-circute the junctions. If the number of ZGLs crossing each junction is large enough, that provides the 2D metallic conductivity (Fig.~\ref{fig7}). According to the weak-localization theory, for delocalization, the characteristic conductance connecting two neighboring sites should be larger than the conductance quantum $2e^2/h$.

The simplest estimation can be done if we assume that the $p$-$n$ junction has a width $L_1\to b\overline{\Delta}/\delta V$ and length $L_2\to b$. Let $|\xi-\xi_c|$ be not much less than 0.5. The number of TI phase inclusions to the OI (or OI to TI) with a size $a$ in this rectangle is $N_{cross}=b^2(\overline{\Delta}/\delta V)|\xi (1-\xi)|/a^2$. Let $a\gg b (\overline{\Delta}/\delta V)$ and $N_{cross}\gg 1$. Then the conductance between two metallic lakes $N_{cross}e^2/h$ exceeds $e^2/h$. Thus, the characteristic conductance, apart from the threshold in $\xi=\xi_c$, is
\begin{eqnarray}\label{sigmajunc2}
\Sigma_{junc}\sim\frac{e^2}{h}\frac{b^2}{a^2}\frac{\overline{\Delta}}{\delta V}|\xi(1-\xi)|.
\end{eqnarray}

In fact, the factor $|\xi_V-\xi_c|=(\overline{\Delta}/\delta V)$ is the proximity of the potential to the threshold. The smallness of this quantity leads to the lake edges fractality. The mean lake size grows like $b|\xi_V-\xi_c|^{-\gamma}$. Replacing $L_2$ by this cluster perimeter, we have a more accurate estimation for
\begin{eqnarray}\label{sigmajunc2}
\Sigma_{junc}\sim\frac{e^2}{h}\frac{b^2}{a^2}\left(\frac{\overline{\Delta}}{\delta V}\right)^{1-\gamma}|\xi(1-\xi)|.
\end{eqnarray}

The total conductivity $\sigma$  is formed upon the series connection of $\Sigma_{junc}$ and the conductivity of p- and n-domains $\sigma_{e,h}\sim e^2/h$ (see Fig.~\ref{fig7}, right):
\begin{eqnarray}\label{sigall}
\frac{1}{\sigma}\approx\frac{1}{\Sigma_{junc}}+\frac{h}{e^2}.
\end{eqnarray}

Hence, $\sigma$ is determined by the lower values of the quantities $\Sigma_{junc}$ and $e^2/h$.

\section{Conclusions and Discussion}
In conclusions, we have qualitatively studied the low-temperature conductivity of the 2D narrow-gap semiconductor with gap and potential fluctuations. Our consideration is based on the random Volkov-Pankratov model (Sec. III).

The finite system without potential fluctuations was considered in Sec. V. In the system, where the gap fluctuations change their sign, internal edge states near zero-gap lines appear. In this section we assumed that electrons move along the zero gap lines neglecting the tunneling. Based on the fact that all ZGLs are finite, we have found that the conductivity of the infinite system tends to zero. The conductance of the square sample is unstable and fluctuates from zero to the conductance quantum. The conductances of wide and long samples near the percolation threshold have been found [Eqs.~(\ref{Sigma3}) and (\ref{Sigma4})]. In a sample, elongated along the electric field, the conductance drops exponentially with the length, while, in the widened sample, it is proportional to the sample width.

We have obtained the expressions for the 2D conductivity of an infinite sample, accounting for the finite edge state width and interedge tunneling [Sec. VI, Eq.~(\ref{cond2})]. The conductivity near permitted bands edges, when the fluctuations of the gap and potential are weak, has been found [Sec. VII, Eq.~(\ref{sig0})].

The system with strong potential fluctuations exceeding the mean gap has been studied (Sec. VIII). These fluctuations result in the decomposition of the sample to $p$- and $n$-domains separated by $p$-$n$ junctions. If the $p$-$n$ junctions are tunneling impenetrable ones, the conductivity between conducting $p$- and $n$-puddles is provided by the edge states crossing $p$-$n$ junctions. The conductivity of such a system was found in the assumption that the edge-states network is dense enough [Eqs.~(\ref{sigmajunc2}) and (\ref{sigall})].

The main assumption of the present paper is the determinative role of the edge states in the system with a fluctuating gap. This differs from \cite{glazman1,glazman2,skinner1,skinner2}, which also provide the existence of $p$ and $n$ puddles connected by tunneling, while, in our case (Sec. VIII), the conductance between puddles is conditioned by the edge states.

Our model of the TI-OI mixture is somewhat reminiscent of the Chalker-Coddington model \cite{chalker-coddington} of integer QHE at a strong potential disorder, where the system is separated on the domains with filling factors $\nu=0$ and $\nu=1$, on the borders of which the 1D chiral channels are formed. This maps the problem onto one of directed links scattering at different nodes. In our case, the lakes of TI or OI are formed due to the HgTe layer width fluctuation. The models are similar in the collisionless motion of carriers along 1D channels. However, in the adiabatic transport model of QHE, electrons move along the equipotential lines with alternating velocity $\propto \nabla V({\bf r})$, while, in our case, the velocity is constant.

Note that narrow-gap semiconductors with strong short-periodic potential fluctuations (see Fig.~\ref{fig1}) can experience the Anderson-Mott transition, which closes the energy gap \cite{girsch}. When the spatial length is large enough, the energy gap can disappear, but the states inside the mean gap are localized; on the contrary, in the layers with the adiabatically fluctuating width and gap considered here, the edge states are formed near the zero gap lines covering the entire sample.

\paragraph*{\bf Acknowledgments.} This research was supported in part by the RFBR, grant No. 17-02-00837.

\end{document}